\documentclass[intlimits,twoside,a4paper]{article}
\usepackage[cp1251]{inputenc}

\usepackage{xcolor}
\usepackage[eqsecnum]{cmpj3}

\usepackage{bm}

\usepackage{amssymb,amsmath}

\issue{2020}{23}{2}{23801}
\doinumber{10.5488/CMP.23.23801}

\title[On the stress overshoot in cluster crystals under shear]%
{On the stress overshoot in cluster crystals under shear%
}
\author[G.P. Shrivastav, G. Kahl]{G.P. Shrivastav, G. Kahl}
\address{Institut f\"ur Theoretische Physik and
  Center for Computational Materials Science (CMS), TU Wien,
  Wiedner~Hauptstra{\ss}e 8-10, A-1040 Wien, Austria}

\date{Received January 31, 2020, in final form March 3, 2020}
\begin{document}

\maketitle

\begin{abstract}
Using non-equilibrium molecular dynamics simulations we study the
yielding behaviour of a model cluster crystal formed by ultrasoft
particles under shear. We investigate the evolution of stress as a
function of strain for different shear rates, $\dot{\gamma}$, and
temperatures. The stress-strain relation displays a pronounced maximum
at the yielding point; the height of this maximum, $\sigma_\text{p}$,
increases via a power law with an increasing shear range and tends to
saturate to a finite value if the limit shear rate goes to zero (at
least within the considered temperature range).  Interestingly, this
behaviour can be captured by the Herschel-Bulkley type model which, for
a given temperature, allows us to predict a static yield stress
$\sigma^{0}_\text{p}$ (in the shear rate tending to zero limit), a
characteristic timescale $\tau_\text{c}$, and the exponent $\alpha$ of the
above-mentioned power-law decay of the $\sigma_\text{p}$ at high shear
rates.
Furthermore, for different temperatures, the $\sigma_\text{p}$ can be
scaled as functions of $\dot{\gamma}$ onto a single master curve when
scaled by corresponding $\tau_\text{c}$ and ${\sigma}_\text{p}^{0}$. Moreover, for a given shear rate, $\sigma_\text{p}$ displays
a logarithmic dependence on temperature. Again, the $\sigma_\text{p}{-}T$ curves for different shear rates can be scaled on a single
logarithmic master curve when scaled by a corresponding fitting
parameters.

\end{abstract}

\section{Introduction}
\label{sec:introduction}

Cluster crystals \cite{mladek2006formation} represent an archetypical
ordered phase encountered in a particular class of soft matter
systems, so-called ultrasoft particles, whose interaction  at
vanishing interparticle distance attains a {\it finite} energy penalty: such
cluster phases can be found in systems if the Fourier transform of the
potentials shows negative components
\cite{likos2001criterion}. At low and
intermediate densities, a disordered phase of clusters of overlapping
particles is found, characterized by a rather polydisperse
distribution in the cluster occupancy. If at a fixed temperature the
density is increased this disordered phase transforms via a first
order phase transition into a cluster BCC phase and then --- upon
further increasing the density --- into a cluster FCC phase. In these
ordered particle configurations, the lattice sites of the BCC and of
the FCC lattice are occupied by clusters of overlapping particles
which are now rather monodisperse in their cluster occupancy
\cite{mladek2006formation,zhang2010reentrant,wilding2013monte,mladek2007clustering,
likos2008cluster,mladek2008multiple}. Even though the fact that mutually repelling
particles are capable of forming stable clusters is at least at first glance
counter-intuitive, the physics behind this intriguing phase is
meanwhile well understood in terms of classical density functional
theory \cite{likos2007ultrasoft,evans2016new}. Furthermore, in the two-dimensional case, the effect of particle-size on the phase diagram of cluster crystals has been explored \cite{caprini2018cluster}. The size of particles is incorporated by introducing a hardcore repulsion in the interaction potential \cite{caprini2018cluster,delfau2016pattern}. In contrast to the fluid cluster-crystal phases where particles inside clusters remain disordered, crystal cluster-crystals are obtained at low temperatures which are characterized by the ordered structure of particles inside a cluster \cite{caprini2018cluster}.

Later contributions in literature have been dedicated to the
equilibrium dynamics of cluster crystals~\cite{mladek2008multiple,nikoubashman2012flow}. Again, these systems
show an intriguingly new behaviour: the longtime dynamics is diffusive
which is the result of the particles hopping from one cluster to a
neighboring cluster, the distribution of the jump length is found to
be exponential at short distances. At large distances, this
distribution follows a power-law decay, i.e., a behaviour reminiscent
of L\'{e}vy flights \cite{coslovich2011hopping}. As expected, the
diffusion coefficient exhibits the Arrhenius behaviour confirming that
the hoping of particles is an activated process. This particular type
of diffusion of particles makes cluster crystals an appropriate model
system to study the mechanical response of defect-rich crystals
\cite{haring2015coarse}, notably the effect of the defect-dynamics on
the yielding of crystalline solids, an area which is less
  explored in the recently developed theories to understand the
  deformation of crystalline solids
  \cite{reddy2020nucleation,nath2018existence}.

Just recently, {\it out-of-equilibrium} investigations have been
dedicated to cluster crystals
\cite{nikoubashman2011cluster,nikoubashman2012flow,nikoubashman2013computer}. In
a recent computer simulation study on the steady-state rheology of
cluster crystals, an intriguing shear-induced self-assembly was
observed, namely a shear-induced fluidization (i.e., string formation)
at very high shear rates. When the shear rate is further
decreased, string-like structures appear and at very low shear rates a
shear shear-banding regime occurs. The shear viscosity for cluster crystals can be obtained in the steady
state via the ratio of the shear stress and the shear rate. In fact,
in a previous computer simulation by Nikoubashman {et al.} \cite{nikoubashman2011cluster} on GEM-8 cluster crystals, it was shown that at a given temperature, the shear viscosity decreases in a power-law manner with an increasing shear rate. In this work, we are focusing on the yielding behaviour of cluster crystals. Therefore, we do not consider the steady state scenario. However, we expect a
qualitatively similar response for a GEM-4 cluster crystal (which is
the focus of the present study).

Furthermore, the dependence of the shear viscosity on the temperature (for a given shear rate) has not been studied systematically in
literature so far. Considering the previously mentioned study by Coslovich { et al.}
\cite{coslovich2011hopping} on the equilibrium diffusion of particles in cluster crystals, we expect that the shear viscosity should decrease with an increasing temperature. In an effort to understand the effect of temperature on the shear viscosity of cluster crystals, a further detailed study is needed.

This contribution is dedicated to the yielding behaviour of cluster
crystals, a feature which, so far, has not been investigated at
all. In particular, we investigate the effect of shear rate and
temperature on the mechanical failure of these materials. In this
work, we study with the help of extensive computer simulations the
yielding behaviour of the cluster crystals under steady shear
conditions by examining the response of the stress as a function of
strain. We will show that the stress exhibits a maximum after an
initial linear increase; the height of this maximum depends on the
shear rate and temperature.  Similar to the majority of the preceding
investigations on cluster forming systems, our investigations are
based on the so-called generalized exponential model interaction with
the index $n$ (GEM-$n$), a potential which combines the bounded nature of
the interaction (required to ensure clustering) with a functionally
simple mathematical form that is easily amenable to computer
simulations. We will show that the stress exhibits a pronounced
maximum after the initial, linear increase. The height of this maximum
depends on the shear rate and on the temperature; in this contribution we
analyse these dependencies via suitably chosen laws. 

The paper is organized as follows: in
section~\ref{sec:model_simulations}, we introduce the model and
details of the simulations and the related protocols. Results are
presented and discussed in section~\ref{sec:results}, while the final
section contains concluding and summarizing remarks and an outlook to
future, related investigations.

\section{Model and simulation details}
\label{sec:model_simulations}

In our cluster crystal system, particles interact via the so-called
generalized exponential (GEM-$n$) potential \cite{mladek2006formation},
assuming in this contribution $n = 4$; this interaction is defined via
\begin{eqnarray}
\label{gem4}
\Phi(r) = \epsilon \exp[-\left(r/d \right)^{4}],
\end{eqnarray}
where $d$ and $\epsilon$ set the length- and energy-scales of the
model, respectively; note that we use --- in contrast to the usual
notation in the previous, related contributions --- the symbol $d$ for the
range of the interparticle interaction, since the conventional symbol
$\sigma$ is reserved in this manuscript to denote the stress.

In the simulations, the GEM-4 potential is truncated and shifted to
zero at a distance $r_\text{c} = 2.2d$. The units of temperature ($T$),
density ($\rho$), and time ($t$), are given by $k_\text{B}T/\epsilon$,
$\rho d^{3}$, and $t_0 = d\sqrt{m/\epsilon}$, respectively. Here, $m$
is the mass of particles and $k_\text{B}$ is Boltzmann's constant. In
our simulations we set the values of $\epsilon$, $d$, $m$, and $k_\text{B}$ equal to unity.

All investigations were carried out with the LAMMPS simulation
package \cite{plimpton1995fast}: we  performed molecular dynamics
(MD) simulations in an NVT ensemble of $N = 3328$ particles (with
periodic boundary conditions), integrating Newton's
equations-of-motions with the Verlet velocity integration scheme
\cite{frenkel2001understanding}; a time increment of $\Delta t = 0.005
t_0$ was used. The temperature was kept constant with the help of
a DPD thermostat which preserves the hydrodynamics
\cite{soddemann2003dissipative}. For each state point, 50 independent
simulation runs were performed; observables were then obtained by
averaging the related data over these runs.

The system was studied at a fixed density, namely $\rho = 6.5$,
and at five different temperatures, i.e., $T = 0.8$, 0.75, 0.7, 0.65,
and 0.6. At these state points, the system is in a stable FCC cluster
phase, where each site of the FCC lattice is occupied by a cluster of
overlapping particles (see phase diagram shown in
\cite{mladek2006formation}). According to the results available in
literature
\cite{mladek2006formation,coslovich2011hopping,mladek2007clustering}
the average number of particles $N_\text{c}$ pertaining to  cluster
ranges for the considered state points around the value $N_\text{c}
\simeq 13$.

\begin{figure}[!t]
\centerline{\includegraphics[width=0.85\textwidth]{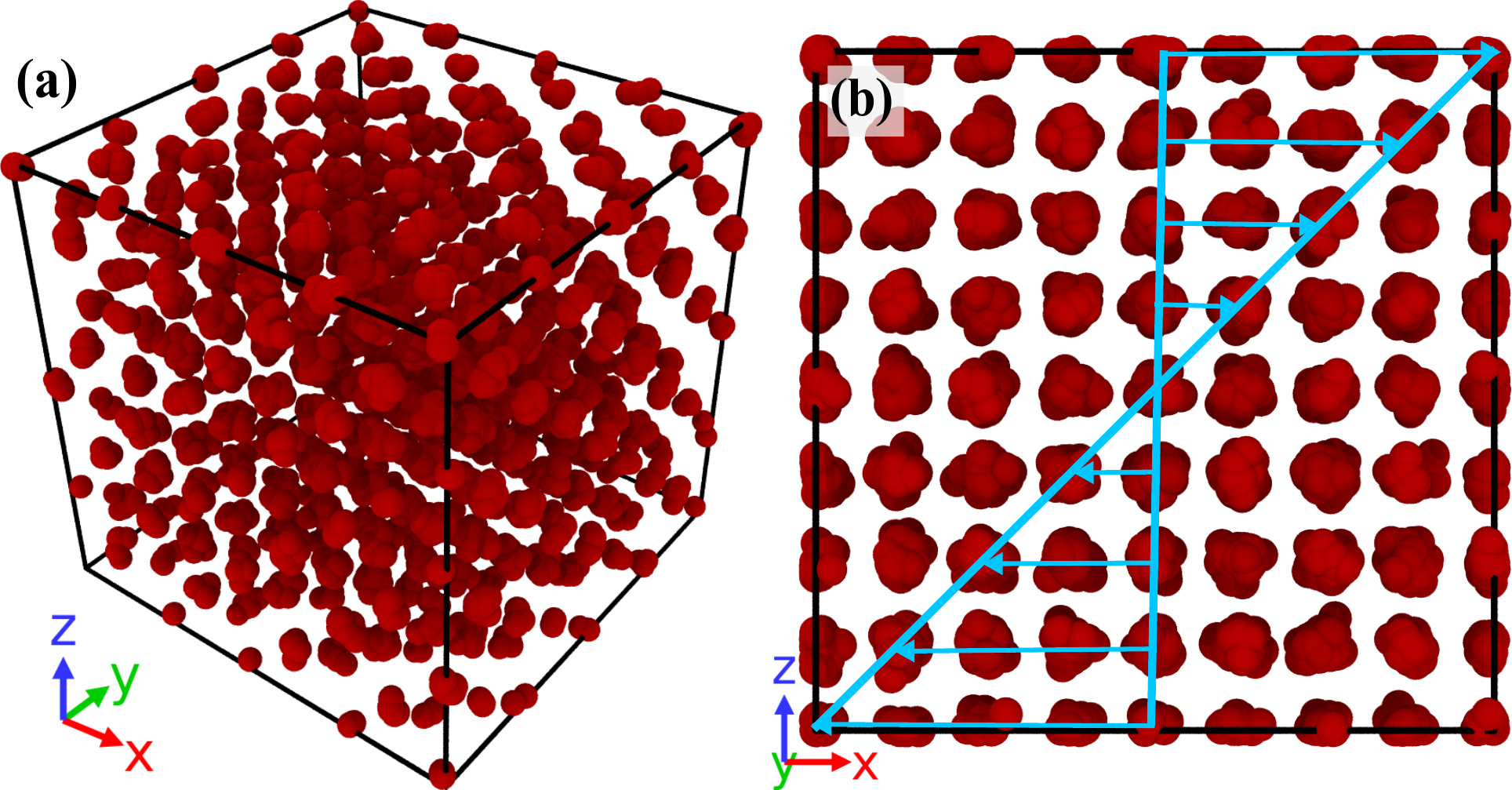}}
\caption{(Colour online) Panel (a): snapshot of an equilibrium configuration cluster
  crystal in the FCC phase at $\rho = 6.5$ and $T = 0.6$ (see text);
  this snapshot was created using the graphic package OVITO
  \cite{stukowski2009visualization}. Panel (b): a schematic view of
  the simulation cell which illustrates how the shear flow is imposed
  in the $(x, z)$-plane on the FCC cluster crystal [using the snapshot
  shown in panel (a)] along the shear direction $x$. The blue arrows
  visualize the velocity gradient along the $z$-direction.}
\label{fig1}
\end{figure}

The initial configurations for our simulations are ideal FCC cluster
crystals where each lattice site is occupied by $N_\text{c} = 13$
completely overlapping particles and assuming a lattice distance that
is compatible with the given value of $\rho$. Starting from this
configuration, the system is equilibrated independently at each of the
above specified temperature values over $10^{6}$ MD steps. A snapshot
of the resulting equilibrium configuration (at $\rho = 6.5$ and $T =
0.6$) is shown in panel (a) of figure~\ref{fig1}.

We shear the bulk cluster crystal by imposing a planar Couette flow in
the $(x, z)$-plane and use Lees-Edwards boundary conditions
\cite{lees1972computer}. To this end, a constant shear rate is applied
along the $x$-direction; thus, in this setup the $z$- and
$y$-directions are the gradient- and vorticity-directions,
respectively. The shear rate, $\dot \gamma$, considered in our
investigations  ranges from, $\dot{\gamma} = 10^{-5}$
to $\dot \gamma = 10^{-1}$ (given in units of $t_0^{-1}$). A schematic
illustration of the setup of our shearing experiment is presented in
panel (b) of figure~\ref{fig1}: here, the front view of the snapshot
(i.e., its projection onto the $(x, z)$-plane) of the equilibrated
cluster crystal (as displayed in panel (a) of figure~\ref{fig1}) is
shown; the velocity gradient is indicated by the blue arrows.

We note that in contrast to \cite{nikoubashman2011cluster}, we
consider here rather low shear rates for the following reasons: (i)~the DPD thermostat (as implemented in the LAMMPS package) becomes in
our investigations unstable at shear rates larger than $\dot{\gamma} =
10^{-1}$; (ii) the present work focuses on an understanding of the
yielding behaviour of cluster crystals in the limit of vanishing shear
rate; 
still it should be mentioned that, on the other
hand, also very low shear rates (i.e., $\dot \gamma \lesssim 10^{-5}$)
are difficult to access, because the related simulations become computationally
expensive.

\section{Results}
\label{sec:results}

In an effort to understand the yielding of the cluster crystal under
shear, we record the evolution of the stress, $\sigma_{xz}(t)$, during
the shearing experiment as a function of strain, $\dot{\gamma}t$,
assuming different values for the shear rate $\dot \gamma$.

From the quantities available during the simulation run, the stress
$\sigma_{xz}(t)$ is calculated via the Irving-Kirkwood expression
\cite{irving1950statistical}:
\begin{eqnarray}
\label{str}
\langle \sigma_{xz}(t)\rangle =
\frac{1}{V}\left\langle\sum_{i}\bigg[mv_{i,x}(t) v_{i,z}(t) +
  \sum_{i>j} r_{ij,x}(t) F_{ij,z}(t) \bigg] \right\rangle.
\end{eqnarray}
Bearing in mind that $m$ and $d$ are set to unity, the unit of
$\langle \sigma_{xz} (t) \rangle$ is $1/t_0^2$. In the above relation,
$V$ represents the total volume of the system, $m$ is the mass of the
particles, $v_{i,x}(t)$ and $v_{i,z}(t)$ represent the $x$- and
$z$-components of the velocity of particle $i$, $r_{ij,x}(t)$ is the
$x$-component of the displacement vector between particles $i$ and
$j$, and $F_{ij,z}(t)$ denotes the $z$-component of the force between
particles $i$ and $j$. The angular brackets in relation (\ref{str})
represent an averaging over the aforementioned 50 independent runs.

In the following subsections we discuss the impact of the shear rate
and of the temperature on the stress.

\subsection{Impact of the shear rate on the stress}
\label{subsec:shear_rate}

\begin{figure}[!t]
\centerline{\includegraphics[width=0.95\textwidth]{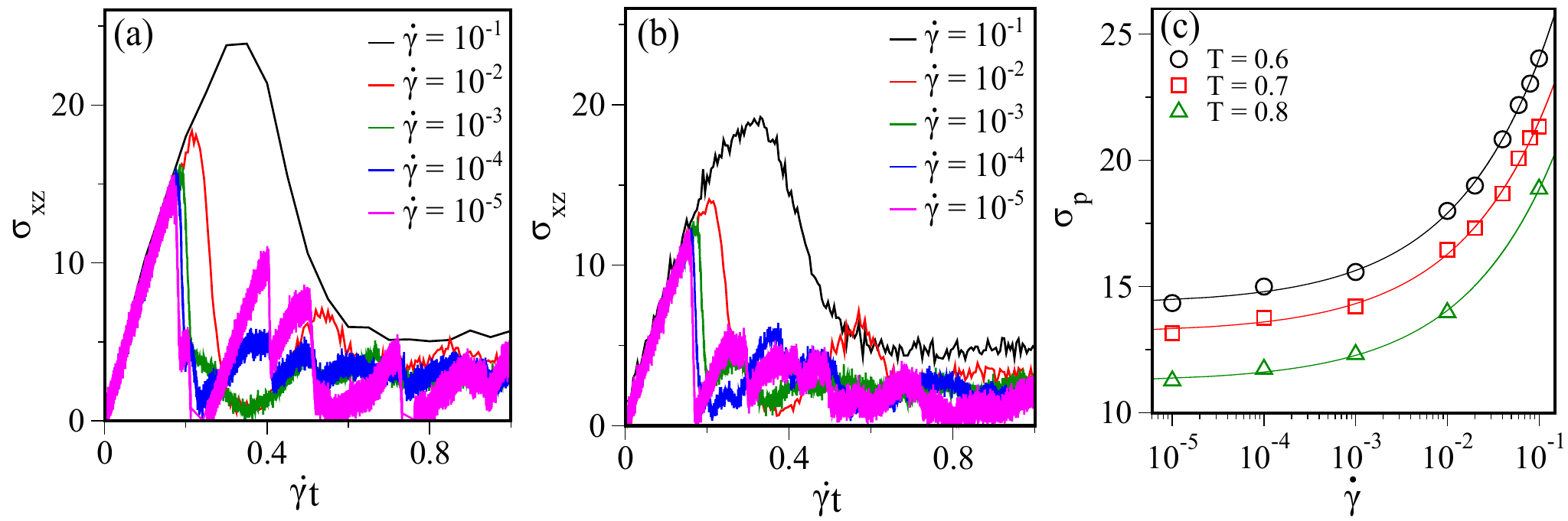}}
\caption{(Colour online) Panels (a) and (b): time evolution of the stress,
  $\langle\sigma_{xz}(t) \rangle$, as a function of strain,
  $\dot{\gamma}t$, for different shear rates, $\dot{\gamma} = 10^{-1},
  10^{-2}, 10^{-3}, 10^{-4}$ and $10^{-5}$ (as labeled), for the case
  $T = 0.6$ and $0.8$, respectively. Panel (c): variation of the stress
  overshoot, $\sigma_\text{p}$ (as defined and discussed in the text),
  as a function of the shear rate, $\dot{\gamma}$, for different
  temperatures (as labeled) in a semi-logarithmic plot. In this panel,
  the solid lines are given by the Herschel-Bulkley type fitting
  function defined in equation~(\ref{hb}).} \label{fig2}
\end{figure}

In panels (a) and (b) of figure~\ref{fig2} we display the time-evolution
of the stress, $\langle \sigma_{xz}(t) \rangle$, as a function of the
strain $\dot{\gamma}t$ for five different shear rates, namely
$\dot{\gamma} = 10^{-1}, 10^{-2}, 10^{-3}, 10^{-4}$, and $10^{-5}$ at
temperatures $T = 0.6$ [panel (a)] and $T = 0.8$ [panel (b)]; similar
graphs also exist for the other temperatures investigated, but they are not
on display here. For all values of $\dot \gamma$ we observe that the
stress first increases with the shear rate and then reaches a
pronounced maximum. At this peak in the stress-strain response curve
the cluster crystal yields and the stress decays beyond this
maximum. We observe that this decay becomes sharper (eventually even
abrupt) as the shear rate is decreased. Beyond this decay, essentially
two different archetypical scenarios of the stress-strain curve can be
identified: (i) for moderate shear rates (i.e., for $\dot \gamma
\simeq 10^{-1}$) the curve levels off without any significant
characteristic features; (ii) for smaller shear rates, however, the
stress-shear curves show characteristic secondary peaks, which become
more and more pronounced as the shear rates decrease.  These
additional peaks can be attributed to partial release events of the
stress via local particle arrangements, a feature which becomes more
pronounced as the shear rate is lowered. When after such a local
rearrangement event the strain increases, the cluster crystal yields
again and another secondary peak appears in the stress-strain curve
which has a lower height than the preceding (and the primary) peak.

The data presented in panels (a) and (b) of figure~\ref{fig2} also
provide evidence that at a fixed temperature, the height of the above
mentioned peak in the stress, denoted henceforward as $\sigma_\text{p}$, decreases as the shear rate is lowered. This also indicates
that the resistance applied by the cluster crystal to the shear forces
reduces as the shear rate is lowered. This observation is visualized
in panel (c) of figure~\ref{fig2}, where we have plotted $\sigma_\text{p}$ as a function of the shear rate $\dot \gamma$ for all the
temperatures investigated. In an effort to quantify this effect and to
provide  more insight into the observed phenomena, we have fitted
$\sigma_\text{p} (\dot{\gamma})$ via the following functional form,
known in literature as the Herschel-Bulkley type expression
\cite{l99,bonn2017yield}, which was originally developed for the
steady state, but which we extend now to the out-of-equilibrium case:
\begin{eqnarray}
\label{hb}
\sigma_\text{p}\left(T, \dot{\gamma}\right) =
\sigma^{0}_\text{p}\left(T\right) + A(T) \dot{\gamma}^{\alpha}.
\end{eqnarray}
The maximum in the stress-strain curve $\sigma_\text{p}$ is also
referred to in literature as the ``static yield stress''
\cite{bonn2017yield}. In case of amorphous solids, it was observed
in computer simulations that the stress overshoot shows a logarithmic
dependence on the shear rate
\cite{varnik2004study,shrivastav2016heterogeneous}. By fitting the
simulation data to the above expression one can extract a value for
$\sigma^{0}_\text{p}$, i.e., the value of the static yield stress as
$\dot{\gamma}$ tends to zero; we conclude that $\sigma^{0}_\text{p}$
will  correspond to the yield stress which is defined as the
minimum stress required to initiate a plastic deformation in the cluster
crystal \cite{bonn2017yield}. We note that alternatively, this
quantity could also be calculated from constant stress simulations
\cite{varnik2004study}, an issue which we postpone to future
investigations. The temperature-dependent parameter $A(T)$ in the
above equation is related to a characteristic timescale, $\tau_\text{c}(T)$, which we  introduce below in equation~(\ref{scl}). Finally,
$\alpha$ is the exponent that characterizes the power-law decay of the
stress overshoot as a function of $\dot \gamma$.

The solid lines in panel (c) of figure~\ref{fig2} show the simulation
data, along with the curves which emerge by fitting these results by
the Herschel-Bulkley type expression defined in equation~(\ref{hb}). The
emerging values of $\sigma^{0}_\text{p}(T)$ and $A(T)$ are listed in
table~\ref{tab1}. The exponent $\alpha$ is found to be essentially
independent of the temperature in the considered range of
temperatures, namely $\alpha \simeq 0.43$. It is worth mentioning that
a related exponent $\alpha$ occurs also in the Herschel-Bulkley model
for the flow curve (i.e., the variation of the steady-state stress as
a function of shear rate). We leave investigations related to this
issue to later contributions.

\begin{table}[!t]
\caption{Parameters $\sigma_\text{p}^{0}$ and $A(T)$ as obtained by
  fitting the simulation data for $\sigma_\text{p}$ via the
  Herschel-Bulkley type expression, given in equation~(\ref{hb}) for the
  five different temperatures investigated in this contribution. The
  temperature-dependent timescale $\tau_\text{c}$ is calculated from
  $A$ and $\sigma_\text{p}$ (as specified in the table).}
\label{tab1}
\vspace{2ex}
\begin{center}
\renewcommand{\arraystretch}{0}
\begin{tabular}{|c||c|c|c|}
\hline
$T$ & $\sigma^{0}_\text{p}$ & $A$ & $\tau_\text{c} =
\left(A/\sigma^{0}_\text{p}\right)^{1/\alpha\strut}$ \\
\hline
0.60 & 14.299 & 26.110 & 4.056  \strut\\
\hline
0.65 & 13.656 & 24.132 & 3.759  \strut\\
\hline
0.70 & 13.170 & 22.526 & 3.484  \strut\\
\hline
0.75 & 12.034 & 22.373 & 4.230  \strut\\
\hline
0.80 & 11.224 & 20.517 & 4.067   \strut\\
\hline
\end{tabular}
\renewcommand{\arraystretch}{1}
\end{center}
\end{table}

\begin{figure}[!t]
\centerline{\includegraphics[width=0.85\textwidth]{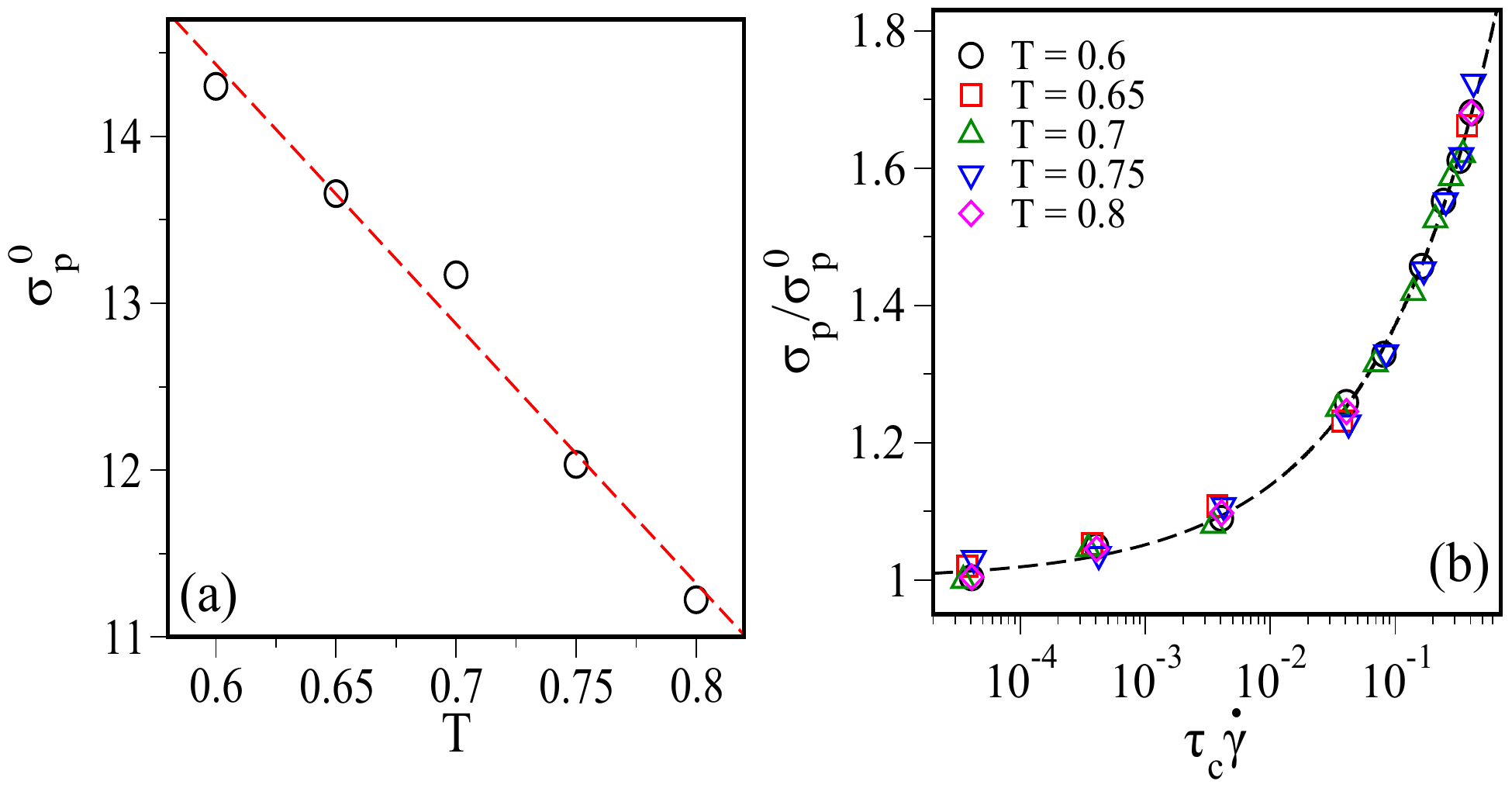}}
\caption{(Colour online) Panel (a): static yield stress, $\sigma^{0}_\text{p}(T)$, as a
  function of temperature $T$, as obtained by fitting the data of
  $\sigma_\text{p}(\dot{\gamma})$ shown in panel (c) of figure~\ref{fig2}
  to the Herschel-Bulkley type expression, equation~(\ref{hb}). Panel (b):
  the symbols show the results of $\sigma_\text{p}/\sigma^{0}_\text{p}$
  (i.e., the data shown in panel (b) of figure~\ref{fig2}) as a function
  of the shear rate, $\dot \gamma$, now scaled by the
  temperature-dependent timescale $\tau_\text{c}$ (see text) for
  different temperatures (as labeled). The black dotted line
  represents the functional form of $\sigma_\text{p} (\dot
  \gamma)/\sigma^{0}_\text{p}$, as given in
  equation~(\ref{scl}).} \label{fig3}
\end{figure}

The static yield stress, $\sigma^{0}_\text{p}(T)$, is displayed as a
function of temperature in panel (a) of figure~\ref{fig3}. This quantity
shows an essentially linear decay with an increasing temperature, a
feature that can probably be attributed to the fact that for a cluster
crystal at equilibrium the number of particles that hop from one
cluster to the neighbouring one decreases with a decreasing temperature.

Further, the different curves $\sigma_\text{p} = \sigma_\text{p}(\dot{\gamma})$, as shown in panel (c) of figure~\ref{fig2} can be
mapped onto one single master curve by scaling the shear rate by a
temperature-dependent timescale, $\tau_\text{c}(T)$, the latter one
being defined as $\tau_\text{c} = (A/\sigma^{0}_\text{p})^{1/\alpha}$; the values of this timescale for different
temperatures are accumulated in  table~\ref{tab1}. Starting from
equation~(\ref{hb}) this master curve has  the following form
\cite{chaudhuri2012inhomogeneous}:
\begin{eqnarray}
\label{scl}
\frac{\sigma_\text{p} (\dot \gamma)}{\sigma^{0}_\text{p}} = 1 +
\left(\tau_\text{c}\dot{\gamma}\right)^{\alpha} .
\end{eqnarray}
In panel (b) of figure~\ref{fig3} we show this master curve onto which
the different $\sigma_\text{p}(\dot \gamma)$-curves can be mapped.
Such a scaling behaviour provides evidence of a universal scenario of
yielding in cluster crystals which --- at least in the considered range
of temperatures --- is independent of the temperature. 

\subsection{The effect of temperature}
\label{subsec:temperature}

\begin{figure}[!t]
\centerline{\includegraphics[width=0.95\textwidth]{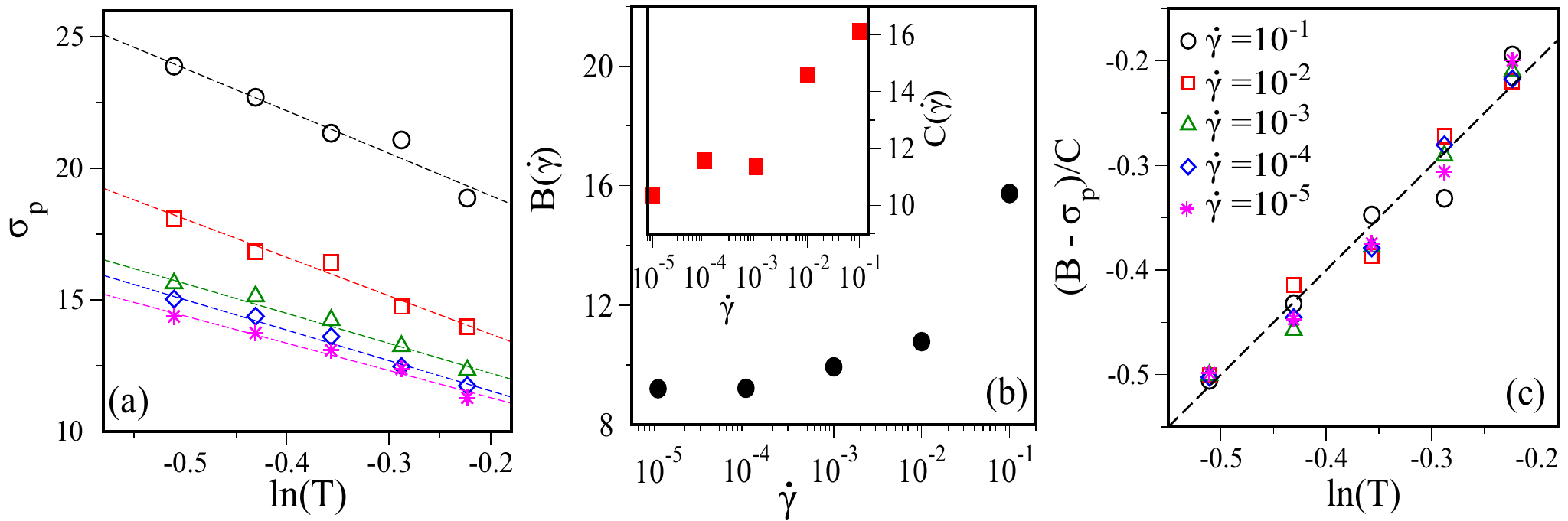}}
\caption{(Colour online) Panel (a): variation of the stress overshoot, $\sigma_\text{p}$, as a function of temperature (in logarithmic units) for shear
  rates, $\dot{\gamma} = 10^{-1}, 10^{-2}, 10^{-3}, 10^{-4}$ and
  $10^{-5}$ [as labeled in panel (c)]. The solid lines originate from
  the logarithimic fittig function, defined in
  equation~(\ref{logfit}). Panel (b): variation of the fitting parameters
  $B\left(\dot{\gamma}\right)$ and $C\left(\dot{\gamma}\right)$ (see
  the inset), introduced in equation~(\ref{logfit}), as functions of the
  shear rate. Panel (c): Scaling of $(B - \sigma_\text{p})/C$ (shown as
  a dashed black line) as a function of the temperatures (on a
  logarithmic scale). The symbols are the related simulation
  data.} \label{fig4}
\end{figure}

It can be seen from panel (c) of figure~\ref{fig2} that the yielding of
cluster crystals strongly depends  on the temperature. In the following,
we analyze this dependence by investigating the behaviour of
$\sigma_\text{p}$ as a function of temperature for different shear
rates. Data shown in panel (a) of figure~\ref{fig4} provide evidence
that $\sigma_\text{p}$ decreases logarithmically as the temperature
increases for all the shear rates considered. Thus, the data can be
fitted via

\begin{eqnarray}
\label{logfit}
\sigma_\text{p}\left(T, \dot{\gamma}\right) =
B\left(\dot{\gamma}\right) - C\left(\dot{\gamma}\right) \ln T,
\end{eqnarray}
where $B\left(\dot{\gamma}\right)$ and $C\left(\dot{\gamma}\right)$
are suitably chosen parameters. The dashed lines in panel (a) of
figure~\ref{fig4} represent the fitting function defined in the above
equation. The related coefficients, $B\left(\dot{\gamma}\right)$ and
$C\left(\dot{\gamma}\right)$, shown in figure~\ref{fig4}~(b), are
themselves functions of the shear rate: (i)
$B\left(\dot{\gamma}\right)$ decreases at high shear rates with $\dot
\gamma$, it saturates at low shear rates; (ii) the $\dot
\gamma$-dependence of $C\left(\dot{\gamma}\right)$ does not show any
significant features. Finally, we point out that the quantity $[B(\dot
  \gamma) - \sigma_\text{p} (T)]/D(\dot \gamma) = \ln T$ can be mapped
onto a single master curve, shown in panel (c) of figure \ref{fig4} as a
black dashed line.


It should be noted that such a logarithmic decay of $\sigma_\text{p}$
as a function of temperature is also observed in thermal glasses
\cite{shrivastav2016heterogeneous,varnik2004study,rottler2005unified}. This
is the more astonishing since such systems are metastable, thus the age
of the sample becomes an important feature.  Cluster crystals, on the
other hand, are equilibrium systems, and aging is in this case not a
relevant phenomenon.


\section{Conclusion and outlook}
\label{sec:conclusion_outlook}

For this contribution we have made an extensive simulation-based study
to understand the impact of the shear rate and of the temperature on
the yielding of cluster crystals which is exposed to shear forces. In
particular, we investigated the behaviour of the stress overshoot for
different shear rates and temperatures by analysing the stress-strain
curve. We have found that for a given temperature the height of the
stress overshoot, $\sigma_\text{p}$, increases in a power-law manner
with an increasing shear rate in a Herschel-Bulkley type manner; the
related exponent is essentially independent of temperature --- at least
for the range of the temperatures considered in this
work. At low shear rates, $\sigma_\text{p}^{0}$ tend to
  saturate to a finite value, $\sigma_\text{p}^{0}$. However, we note
  here that a systematic finite-size analysis is needed to ensure the
  non-vanishing value of $\sigma_\text{p}^{0}$. Further, the behaviour
of the stress overshoot as a function of shear rate is found to have a
universal nature, thus different curves of $\sigma_\text{p}(\dot{\gamma})$ can be mapped onto a single master curve. The
static yield stress is found to decay in a linear manner with
temperature. At a given shear rate, the height of the stress overshoot
exhibits a logarithmic dependence on the temperature.  This behaviour
is also found to be universal, thus different $\sigma_\text{p}(T)$
curves can be mapped onto a single master curve. We note that this
logarithmic behaviour is a characteristic feature of defect-rich
systems such as thermal glasses
\cite{varnik2004study,shrivastav2016heterogeneous}, as also predicted
by the Ree-Eyring viscosity theory \cite{l99,eyring1936viscosity}.

Further, a systematic understanding of the mechanical response of cluster crystals is still missing. A complete calculation of the various elastic constants would be desirable but it requires expensive computer simulations. The bulk modulus of cluster crystals, studied via density functional theory and Monte Carlo simulations was discussed by Mladek { et al.} \cite{mladek2007phase}. In later years, these results
for the bulk modulus were compared with the related data obtained from
the dispersion relations for cluster crystals by H\"{a}ring { et al.} \cite{haring2015coarse}. 

\section{Acknowledgements}

The authors acknowledge financial support by the Austrian Science
Foundation (FWF) under Proj. No. I3846. The computational results
presented have been achieved using the Vienna Scientific Cluster
(VSC). GK would like to express his gratitude towards Ihor Mryglod for
so many years of strong personal and scientific ties.

\newpage
\ukrainianpart

\title{Про викид напруження в кластерних кристалах під дією зсуву}
\author{Г.П. Шрівастав, Г. Каль}
\address{Інститут теоретичної фізики і  центр обчислювального матеріалознавства
	 (CMS), TU Відень,
   A-1040 Відень, Австрія}

\makeukrtitle 

\begin{abstract}
Використовуючи симуляції методом нерівноважної молекулярної динаміки, ми досліджуємо 
 поведінку податливості	модельного кластерного кристалу, сформованого ультрам'якими частинками під дією зсуву. Ми досліджуємо еволюцію напруження як функцію деформації для різних швидкостей зсуву, $\dot{\gamma}$, і температур.  Відношення напруження-деформація
показує  чіткий максимум в точці податливості; висота цього максимума, $\sigma_\text{p}$,
зростає за степеневим законом зі збільшенням області зсуву і прямує до насичення до кінцевого значення, коли гранична швидкість зсуву прямує до нуля (принаймні в межах розглянутого діапазону температур). Виявилося, що цю поведінку можна отримати   з допомогою моделі Гершеля-Балклі, яка при заданій температурі дозволяє зробити передбачення  для статичного податливого напруження 
$\sigma^{0}_\text{p}$ (при швидкості зсуву, що прямує до нуля), характерного часового масштабу $\tau_\text{c}$ та степеня $\alpha$ вище-згаданого степенового закону
 $\sigma_\text{p}$  при високих швидкостях зсуву.
 До того ж, при різних температурах,   $\sigma_\text{p}$ можуть бути промасштабовані як функції  $\dot{\gamma}$  в єдину криву при масштабуванні відповідними значеннями $\tau_\text{c}$ і ${\sigma}_\text{p}^{0}$.  Крім того, при заданій швидкості зсуву  $\sigma_\text{p}$ показує логарифмічну залежність від температури. Знову, криві $\sigma_\text{p}{-}T$  для різних швидкостей зсуву  можуть бути промасштабовані в єдину логарифмічну криву при масштабуванні відповідними підгоночними параметрами.
\keywords реологія, кластерний кристал, податливість, молекулярна динаміка, перенапруження
\end{abstract}
\lastpage
\end{document}